\documentclass{emulateapj}

\usepackage{epsfig}

\newcommand\rms{{\rm s}} 
\def\rmp{{\rm p}}
\newcommand\rmd{{\rm d}} 
\newcommand\rme{{\rm e}} 
\newcommand\itq{{\it q}}

\newcommand\RR{{\rm RR}} 
\newcommand\DR{{\rm DR}}

\shortauthors{N. R. Badnell}
\shorttitle{DR of Fe $3\rmp^q$ ions}

\begin{document}

\title{Dielectronic recombination of F\rme $\,\,3\rmp^\itq$ ions: a key ingredient for describing
X-ray absorption in active galactic nuclei}

\author{\sc  N. R. Badnell}

\affil{Department of Physics, University of Strathclyde, Glasgow G4 0NG, UK}

\begin{abstract}
We have carried-out multi-configuration Breit--Pauli {\sc autostructure} calculations for the
dielectronic recombination (DR) of Fe$^{8+}$ -- Fe$^{12+}$ ions. We obtain total DR rate coefficients
for the initial ground-level which are an order of magnitude larger than those corresponding to
radiative recombination (RR), at temperatures where Fe $3\rmp^q$ ($q=2-6$) ions are abundant in 
photoionized plasmas. The resultant total (DR+RR) rate coefficients are then an order of magnitude
larger than those currently in use by photoionized plasma modeling codes such as {\sc cloudy, ion}
and {\sc xstar}. These rate coefficients, together with our previous results for $q=0$ and $1$, 
are critical for determining the ionization balance of the M-shell Fe ions which give rise to
the prominent unresolved-transition-array X-ray absorption feature found in the spectrum of
many active galactic nuclei. This feature is poorly described by {\sc cloudy} and {\sc ion},
necessitating an {\it ad hoc} modification to the low-temperature DR rate coefficients.
Such modifications are no longer necessary and a rigorous approach to such modeling
can now take place using these data.
\end{abstract}


\section{Introduction}
The iron M-shell ions (predominantly Fe$^{7+}$ -- Fe$^{13+}$) give rise to strong X-ray absorption lines due to
$n=2-3$ electronic inner-shell transitions. These are seen in the spectra of active galactic nuclei (AGNs)
observed by {\it Chandra} and {\it XMM-Newton} \citep{Sak01} as an unresolved-transition-array (UTA). 
The shape of the UTA feature can be used as a powerful diagnostic --- see \cite{Beh01} for a detailed discussion,
and whose atomic database for describing it has been updated recently by \cite{Gu06}. 
However, \cite{Net03} have pointed-out problems in modeling this shape and so \cite{Net04}, using {\sc ion},
and \cite{Kra04}, using {\sc cloudy}, have suggested increasing the magnitude of the recombination rate 
coefficients for Fe ions, particularly the $3\rmp^q$ ($q=1-6$)
ions Fe$^{13+}$ -- Fe$^{8+}$, by postulating {\it ad hoc} low-temperature dielectronic recombination (DR)
rate coefficients. The net effect is to shift the ionization balance towards the neutral end. This
brings the modeling results towards accord with observation, but it is not rigorous since there is
little to constrain the DR rate coefficients which they use.

The extant recombination data is effectively all radiative (RR) at photoionized plasma temperatures
\footnote{When we refer to photoionized or electron-collisional plasma temperatures we mean the
temperatures at which Fe $3\rmp^q$ ions are abundant in such plasmas --- see \cite{Kall01} ({\sc xstar})
and \cite{Mazz98}, respectively.}.
The recommended DR data
for these Fe ions \citep{Arna92a} is based upon the high-temperature electron-collisional plasma results
of \cite{Jac77}, which give essentially zero-contribution from DR at photoionized plasma temperatures.
There is much evidence from the Fe L-shell, both experimental and theoretical \citep[see, e.g.,][]{Sav06},
that a significant contribution can be expected from DR at low temperatures, due principally to `non-dipole' 
core-excitations which were not (needed to be) considered by \cite{Jac77}.

Recently, measurements by \cite{Sch06}
\footnote{\cite{Sch06} also give an up-to-date list of references of observations of the AGN X-ray absorption feature.}
 on Fe$^{13+}$ ($q=1$) at the Heidelberg  heavy-ion test storage ring have 
found the contribution from DR to the total recombination rate coefficient to be an order of magnitude 
larger than that due to RR, at photoionized  plasma temperatures. We have carried-out a detailed theoretical 
analysis of these experimental results \citep{Bad06_Fe13}. We found some disagreements, especially at energies 
which affect the DR rate coefficient at photoionized  plasma temperatures. Nevertheless, our DR rate
coefficient for Fe$^{13+}$ is also an order of magnitude larger than our RR one, but it is up to a third smaller
than the experimentally-based one over $10^4 - 10^5$~K. We now report-on the results of calculations for the
computationally demanding Fe $3\rmp^q$  ($q=2-6$) ions.

The remainder of the paper is organized as follows: in Section 2 we describe our methodology, 
in Section 3 we present our results and make comparisons with the results of earlier works, and then
make some concluding remarks.

\section{Methodology}
Our recent comprehensive study \citep{Bad06_Fe13} of the DR of Fe$^{13+}$ ($q=1$), including
detailed comparisons of our theoretical cross sections with those of the high-energy resolution 
measurements of \cite{Sch06}, provides us with a guide for our approach to the 
$3\rmp^q$ ($q=2-6$) ions. These are
both computationally more demanding and there are no (published) experimental results for them.
We give the essentials of the new work below and refer the reader to \cite{Bad06_Fe13} for 
a more detailed exposition of the atomic physics.
But, we note in particular that a level-resolved treatment is critical for the determination of 
accurate  DR rate coefficients at photoionized plasma temperatures, but the results of such calculations
are sparse for the M-shell.

\subsection{Theory}
The total dielectronic recombination rate coefficient, $\alpha^\DR_{\nu} (T)$, 
from an initial state $\nu$ of an $N$-electron ion is given, at a temperature $T$, by \citep{Bur64}
\begin{eqnarray}
\alpha^\DR_{\nu}(T)&=&\left({4\pi a^2_0 I_{\rm H} \over k_{\rm B} T}\right)^{3/2}
\sum_j{\omega_{j} 
\over 2\omega_{\nu}}\,{\rm e}^{-E_{\rm c}/(k_{\rm B} T)}\nonumber \\
 &\times&{ \sum_{i,l}A^{{\rm a}}_{j \rightarrow \nu, E_cl} \, A^{{\rm r}}_{j \rightarrow i}
\over \sum_{h} A^{{\rm r}}_{j \rightarrow h} + \sum_{m,l} A^{{\rm a}}_{j \rightarrow m, E_cl}}\, ,
\label{eq12}
\end{eqnarray}
where $\omega_j$ is the statistical weight of the
$(N+1)$-electron doubly-excited resonance state $j$, $\omega_\nu$ is the statistical weight
of the initial state and the autoionization ($A^{\rm a}$) and radiative
($A^{\rm r}$) rates are in inverse seconds. Here, $E_{\rm c}$ is the energy of the continuum 
electron (with orbital angular momentum $l$), which is fixed by the position of the resonances, 
and $I_{\rm H}$ is the ionization potential energy of the hydrogen atom (both in the
same units of energy), $k_{\rm B}$ is the Boltzman constant, $T$ the electron temperature
and $(4\pi a^2_0)^{3/2}=6.6011\times10^{-24}$ cm$^3$.

We have used {\sc autostructure} \citep{Badn86a} to carry-out multi-configuration Breit--Pauli 
calculations of all of the necessary rates and energies, as detailed next. 

\subsection{The F\rme $\,\,3\rmp^\itq$ target}
We describe the $N$-electron target by the following configurations (assuming a closed Ne-like core):

\( \begin{array}{lll}
& & \\
1: \quad 3\rms^2 3\rmp^q\,,   & 2: \quad 3\rms 3\rmp^{q+1}\,,   & 3: \quad 3\rms^2 3\rmp^{q-1} 3\rmd\,,\\
& & \\
4: \quad 3\rmp^{q+2}\,, & 5: \quad 3\rms 3\rmp^q 3\rmd\,,  & 6: \quad 3\rms^2 3\rmp^{q-2} 3\rmd^2\, \\
& & \\
7: \quad 3\rmp^{q+1} 3\rmd\,. & & \\
& &
\end{array} \)

\noindent
(If a given value of $q$ results in an occupation number $<0$ or $>6$ then that `configuration'
does not exist for said ion.)
This target expansion allows for both $3\rms$ and $3\rmp$ $\Delta n=0$ sub-shell promotions from the ground
configuration, as well as including important interacting configurations. We denote it as `7CF'.
We also investigated the accuracy of this representation for DR by carrying-out a further
calculation for each ion which included the additional configuration
interaction due to

\( \begin{array}{l}
\\
8: \quad 3\rms 3\rmp^{q-1} 3\rmd^2\, . \\
\\
\end{array}\)

\noindent
We denote the combined set as `8CF'.

The contribution from higher energy ($\Delta n=1$) promotions has no effect at photoionized
plasma temperatures, while in electron-collisional plasmas we expect such contributions
to be less than 10\% of the $\Delta n=0$, following our results for $q=1$ \citep{Bad06_Fe13}.
(The relative importance of $\Delta n=1$ contributions decreases as $q$ increases. This is because 
the inner-shell $2-3$ contribution is further suppressed by additional core re-arrangement 
autoionizing transitions and the $3-4$ outer-shell contribution peaks at a temperature which
is closer to that of the $\Delta n=0$ peak.)

All relevant $(N+1)$-electron autoionization and radiative rates are then determined, for the
given target expansion, for all Rydberg states up to $n=1000$ and $l=9-13$, depending on $q$, 
and the total DR rate coefficient is then determined according to equation (1). 

In addition, we use observed target energies \citep{NIST} wherever possible. This minimizes
the sensitivity to the Maxwellian exponential factor at low temperatures, which is critical for 
application to photoionized plasmas. Since the observed energies are incomplete, 
we adopt the expedient strategy of using calculated level splittings to adjust the position of
missing levels of a term, relative to an observed one. Similarly, we adjust (unobserved) terms to
maintain the splitting with observed ones of the same symmetry. Finally, for higher energy configurations
(typically, configuration number 4 and above) where there are no observed level energies, we adjust
the entire configuration position by consideration of the average shift of lower-lying configurations.

Radiative recombination may be expected to be important at photoionized plasma temperatures
and so we have also calculated total RR rate coefficients, $\alpha^\RR_\nu (T)$, with 
{\sc autostructure}, following \cite{Bad06_RR}.

\subsection{Fits}
For ease of use, we fit our total recombination rate coefficients to the usual
functional forms: for DR,
\begin{eqnarray}
\alpha^\DR_\nu(T)&=&T^{-3/2}\sum_i c_i \exp\left({-E_i / T}\right)\,,
\label{DRfit}
\end{eqnarray}
where the $E_i$ are in the units of temperature, $T$ (K), and the units of $c_i$ are 
then cm$^3 \rms^{-1}$~K$^{3/2}$; for RR,
\begin{eqnarray}
\label{fit1}
&\alpha&^\RR_\nu(T)=A  \\
&\times& \left[\sqrt{T/T_0}\left(1+\sqrt{T/T_0}\right)^{1-B}\left(1+\sqrt{T/T_1}\right)^{1+B}\right]^{-1},
\nonumber
\end{eqnarray}
where, for greater accuracy, $B$ may be replaced as
\begin{equation}
\label{fit2}
B \rightarrow B + C \exp(-T_2/T ).
\end{equation}
Here, $T_{0,1,2}$ are in units of temperature (K), the units of $A$ are  cm$^3$~s$^{-1}$, while $B$ and $C$ 
are dimensionless. 

\section{Results}
In Figures 1 -- 5 we present our recombination rate coefficients for the initial ground-level
of Fe$^{12+}$ -- Fe$^{8+}$ ($q=2-6$) and compare
them with the DR rate coefficients recommended by \cite{Arna92a}.
The data of \cite{Arna92a} are based principally upon the 
results of \cite{Jac77}, but include an estimate of the contribution from $2\rmp - 3\rmd$ inner-shell
transitions as well --- these were not included by \cite{Jac77}. The work of \cite{Jac77} was for
application to high-temperature electron-collisional plasmas and their results have little in the
way of a low-temperature contribution --- a modest one may just be seen for $q=4$ and 5 at around $10^5$~K.

The main result, common to all ions, is the order of magnitude difference between the DR and RR
rate coefficients at photoionized plasma temperatures. This is similar to the situation
found for $q=1$ \citep{Sch06, Bad06_Fe13}. Thus, the {\it ad hoc} modifications proposed by \cite{Net04}
and by \cite{Kra04} were conservative --- increasing the total recombination rate coefficient by
factors of only 2--4. These new rate coefficients 
(plus $q=0,1$ data)
can be expected to have a significant effect upon the
ionization balance of Fe$^{7+}$ -- Fe$^{13+}$ in photoionized plasmas and, in turn, the modeling of the
UTA X-ray absorption feature in AGNs.
On comparing our 7CF and 8CF results, we note also that sensitivity to near threshold resonance positions 
at low temperatures does not appear 
to become an issue until $\lesssim 10^4$~K, i.e., below the main temperatures of interest for
photoionized plasmas.

At electron-collisional plasma temperatures we find reasonable agreement with the recommended data
of \cite{Arna92a}. The strength of the contribution from `low-temperature' resonances extends to
enhancing the high-temperature peak in many cases. Even at $10^7$~K, which is far-off equilibrium,
our present results are still (a little) larger than those recommended by \cite{Arna92a}, except for the case of $q=2$.
Even in this case ($q=2$) we expect the $\Delta n=1$ contribution to result-in no more than a 10\%
increase at a few times $10^6$~K, rising to at most 20\% by $10^7$~K, and less for lower charge ions,
based-upon our results for $q=1$ \citep{Bad06_Fe13} and our comments in Section 2.2. 
Thus the present results can, and should, be used for modeling electron-collisional 
plasmas as well.

In Tables 1 and 2 we present our fit coefficients for these DR and RR rate coefficients, respectively, as 
defined by Equations (2) and (3). 
For convenience, we include the results for $q=1$ \citep{Bad06_Fe13} and
$q=0$ \citep{Alt06} as well. Results for all higher charge states may be found online \citep{Bad06_DR_web, Bad06_RR_web},
following the work of \cite{Bad03a} and \cite{Bad06_RR} for DR and RR, respectively.

\section{Concluding Remarks}
We have reported new DR rate coefficients for Fe $3\rmp^q$ ($q=2-6$) ions which give rise to
total recombination rate coefficients which are an order of magnitude larger, at photoionized
plasma temperatures, than those currently recommended \citep{Arna92a} and routinely used by
modeling codes such as {\sc cloudy, ion} and {\sc xstar}. These new rate coefficients can be
expected to change significantly the ionization balance of the Fe M-shell ions which give rise to the
important UTA X-ray absorption feature seen in the spectra of AGNs observed by {\it Chandra} and
{\it XMM-Newton}, and to enable rigorous modeling of it to be carried-out now using these data.

Similarly, large low-temperature DR contributions can be expected for $3\rmp^q$ ions of other 
elements of astrophysical importance, e.g., Si, S, Ar, Ni. A move into the $3\rmd$ sub-shell of lower-charge Fe 
ions, using a level-resolved approach, is also desirable.

\acknowledgements

This work was supported in part by PPARC Grant No. PPA$\backslash$G$\backslash$S2003$\backslash$00055
with the University of Strathclyde.


\clearpage

\begin{deluxetable}{rllllllll}
\tablecaption{DR fitting coefficients, $c_i$~({\rm cm}$^3$\rms$^{-1}$K$^{3/2}$) and $E_i$(K), for the initial 
ground-level of {\rm Fe} $3\rmp^q$ ($q=0-6$) ions.\label{tab1}
}
\tablewidth{0pt}
\tablehead{
\colhead{$q$}&\colhead{$c_1$}&\colhead{$c_2$}&\colhead{$c_3$}&\colhead{$c_4$}&\colhead{$c_5$}&
\colhead{$c_6$}&\colhead{$c_7$}&\colhead{$c_8$}
}
\startdata
0 & 5.636($-$4)\tablenotemark{a} & 7.390($-$3) & 3.635($-$2) & 1.693($-$1) & 3.315($-$2) & 2.288($-$1) & 7.316($-$2) \\
1 & 1.090($-$3) & 7.801($-$3) & 1.132($-$2) & 4.740($-$2) & 1.990($-$1) & 3.379($-$2) & 1.140($-$1) & 1.250($-$1) \\
2 & 3.266($-$3) & 7.637($-$3) & 1.005($-$2) & 2.527($-$2) & 6.389($-$2) & 1.564($-$1) \\
3 & 1.074($-$3) & 6.080($-$3) & 1.887($-$2) & 2.540($-$2) & 7.580($-$2) & 2.773($-$1) \\
4 & 9.073($-$4) & 3.777($-$3) & 1.027($-$2) & 3.321($-$2) & 8.529($-$2) & 2.778($-$1) \\
5 & 5.335($-$4) & 1.827($-$3) & 4.851($-$3) & 2.710($-$2) & 8.226($-$2) & 3.147($-$1) \\
6 & 7.421($-$4) & 2.526($-$3) & 4.605($-$3) & 1.489($-$2) & 5.891($-$2) & 2.318($-$1) \\
\noalign{\vskip 1.5ex}
\hline
\noalign{\vskip .8ex}

$q$& $E_1$ & $E_2$ & $E_3$ & $E_4$ & $E_5$ & $E_6$ & $E_7$ & $E_8$ \\

\noalign{\vskip .8ex}
\hline
\noalign{\vskip .8ex}
0 & 3.628(3) & 2.432(4) & 1.226(5) & 4.351(5) & 1.411(6) & 6.589(6) & 1.030(7) \\
1 & 1.246(3) & 1.063(4) & 4.719(4) & 1.952(5) & 5.637(5) & 2.248(6) & 7.202(6) & 3.999(9) \\
2 & 1.242(3) & 1.001(4) & 4.466(4) & 1.497(5) & 3.919(5) & 6.853(5) \\
3 & 1.387(3) & 1.048(4) & 3.955(4) & 1.461(5) & 4.010(5) & 7.208(5) \\
4 & 1.525(3) & 1.071(4) & 4.033(4) & 1.564(5) & 4.196(5) & 7.580(5) \\
5 & 2.032(3) & 1.018(4) & 4.638(4) & 1.698(5) & 4.499(5) & 7.880(5) \\
6 & 3.468(3) & 1.353(4) & 3.690(4) & 1.957(5) & 4.630(5) & 8.202(5)
\enddata
\tablenotetext{a}{Note, (n) denotes $\times 10^n$.}
\end{deluxetable}

\begin{deluxetable}{rllllll}
\tablecaption{RR fitting coefficients for the initial ground-level of {\rm Fe} $3\rmp^q$ ($q=0-6$) ions.\label{tab2}
}
\tablewidth{0pt}
\tablehead{
  \colhead{$q$} &    \colhead{$A$~(cm$^3$s$^{-1}$)}     &   \colhead{$B$}    &    \colhead{$T_0$(K)}
     &    \colhead{$T_1$(K)}     &   \colhead{$C$}    &    \colhead{$T_2$(K)}      }
\startdata
  0 & 1.179($-$9) & 0.7096 & 4.508(2) & 3.393(7) & 0.0154 & 3.977(6) \\
  1 & 1.050($-$9) & 0.6939 & 4.568(2) & 3.987(7) & 0.0066 & 5.451(5) \\
  2 & 9.832($-$10)& 0.7146 & 3.597(2) & 3.808(7) & 0.0045 & 3.952(5) \\
  3 & 8.303($-$10)& 0.7156 & 3.531(2) & 3.554(7) & 0.0132 & 2.951(5) \\
  4 & 1.052($-$9) & 0.7370 & 1.639(2) & 2.924(7) & 0.0224 & 4.291(5) \\
  5 & 1.338($-$9) & 0.7495 & 7.242(1) & 2.453(7) & 0.0404 & 4.199(5) \\
  6 & 1.263($-$9) & 0.7532 & 5.209(1) & 2.169(7) & 0.0421 & 2.917(5)
\enddata
\end{deluxetable}

\clearpage

\begin{figure}
  \begin{center}
  \epsfig{file=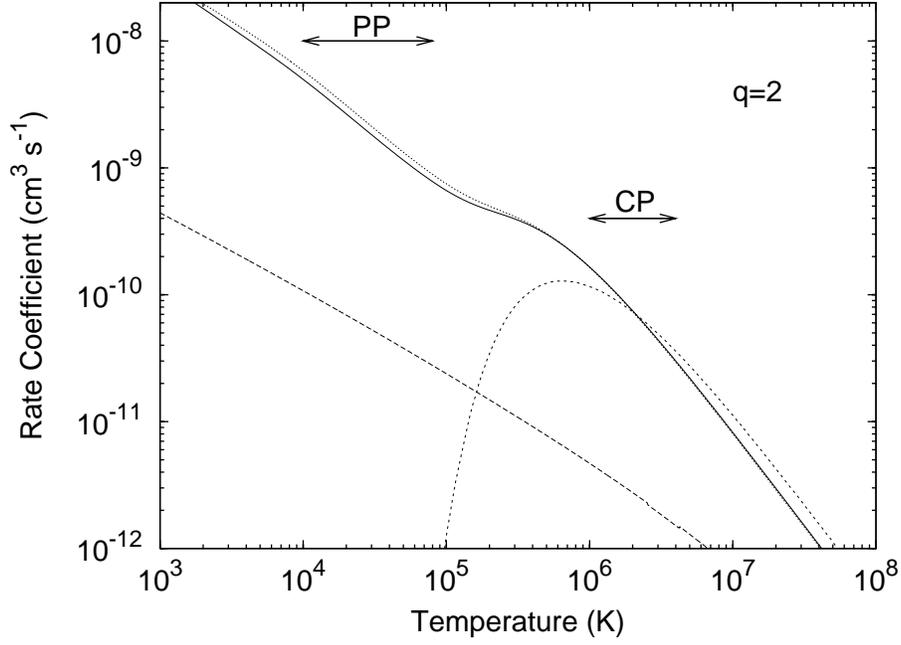}
  \end{center}
  \caption[]{Total ground-level rate coefficients for Fe$^{12+}$ ($q=2$).  Solid curve, DR (7CF); dotted curve, 
DR (8CF); long-dashed curve, RR;  all present {\sc autostructure} results. Short-dashed curve, recommended DR data
of \cite{Arna92a}. PP and CP denote typical photoionized and electron-collisional
plasma temperature ranges, respectively, for Fe$^{12+}$ \citep{Kall01} and \citep{Mazz98}. }
  \label{Fig_q2}
\end{figure}

\begin{figure}
  \begin{center}
  \epsfig{file=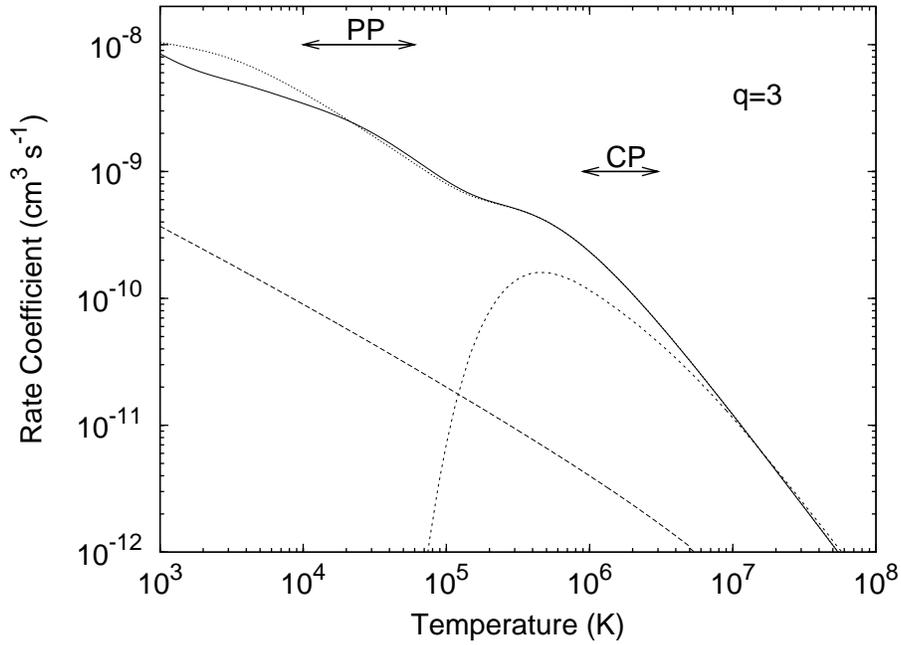}
  \end{center}
  \caption[]{As Figure 1, but for Fe$^{11+}$ ($q=3$).}
  \label{Fig_q3}
\end{figure}

\begin{figure}
  \begin{center}
  \epsfig{file=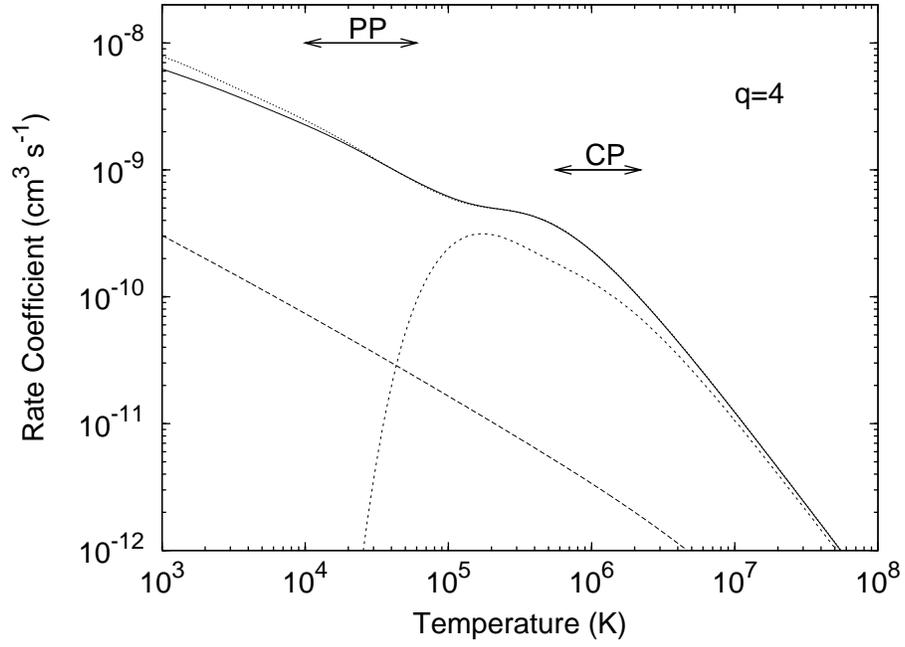}
  \end{center}
  \caption[]{As Figure 1, but for Fe$^{10+}$ ($q=4$).}
  \label{Fig_q4}
\end{figure}

\begin{figure}
  \begin{center}
  \epsfig{file=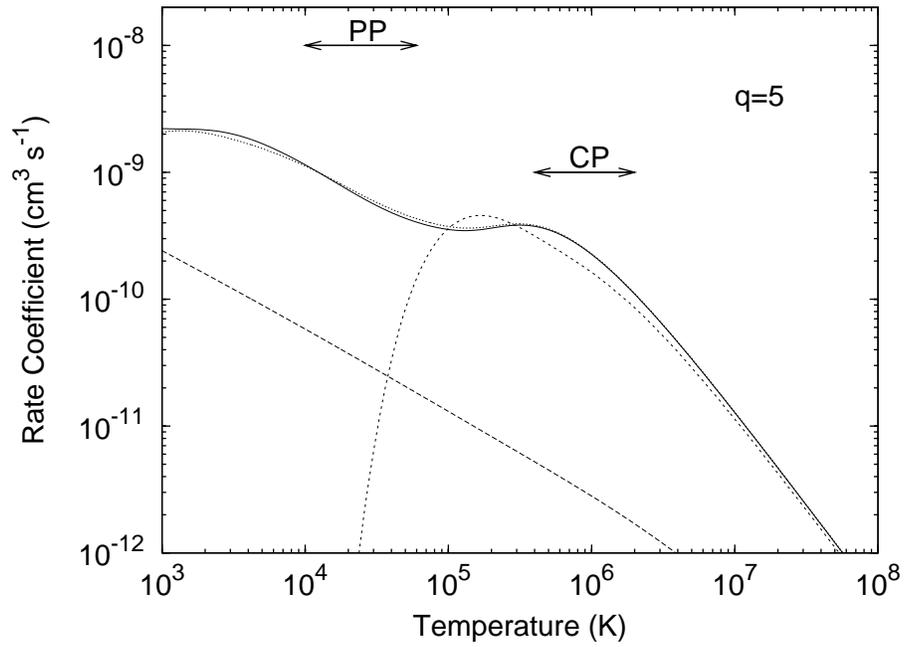}
  \end{center}
  \caption[]{As Figure 1, but for Fe$^{9+}$ ($q=5$).}
  \label{Fig_q5}
\end{figure}

\begin{figure}
  \begin{center}
  \epsfig{file=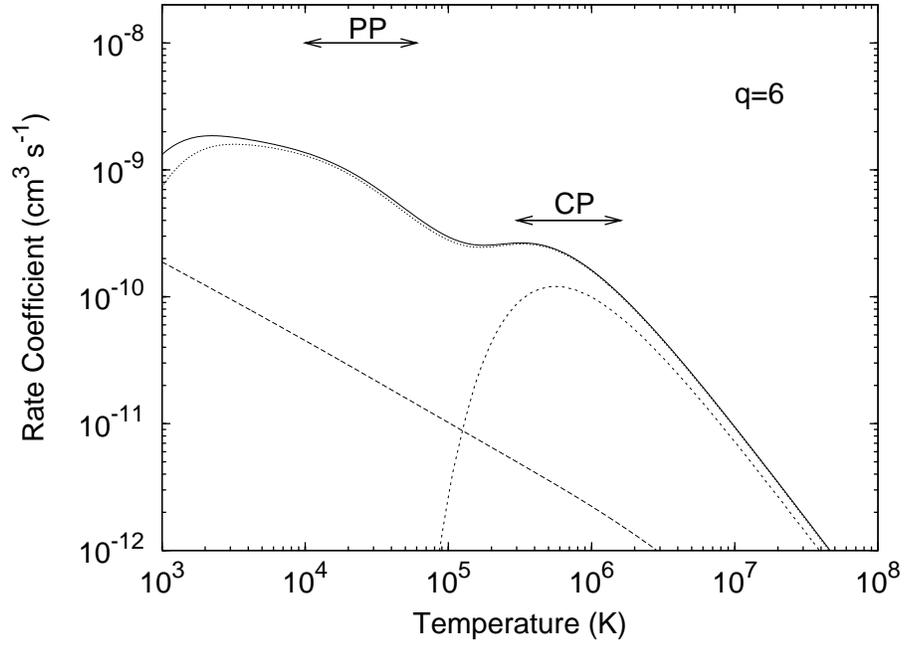}
  \end{center}
  \caption[]{As Figure 1, but for Fe$^{8+}$ ($q=6$).}
  \label{Fig_q6}
\end{figure}

\end{document}